\documentclass{llncs}

\usepackage{amssymb}
\usepackage{graphics,color}      % usual driver
\usepackage{verbatim}
\usepackage{epsfig}
\usepackage{amsmath}    % need for subequations
\usepackage{amsfonts}
\usepackage{theorem}
\usepackage{times}

\newenvironment{prevproof}[2]{\noindent {\em {Proof of
{#1}~\ref{#2}:}}}{$\blacksquare$\vskip \belowdisplayskip}

\newcommand{\N}{\mathbb{N}}

\newcommand{\E}{\mathbb{E}}

\newcommand{\GSnote}[1]{{\textbf{ } }}
\newcommand{\GVnote}[1]{{\textbf{ } }}
\newcommand{\Pnote}[1]{{\textbf{ } }}
\newcommand{\Cnote}[1]{{\textbf{ } }}

\newtheorem{fact}       [theorem]   {Fact}
\def\G{{\mathcal{G}}}

\title{On the Complexity of Nash Equilibria of Action-Graph Games}

\author{
Constantinos Daskalakis\thanks{University of California, Berkeley.
Email: {\tt costis@cs.berkeley.edu}.} \and Grant
Schoenebeck\thanks{University of California, Berkeley. Email: {\tt
grant@eecs.berkeley.edu}.} \and Gregory Valiant\thanks{University of
California, Berkeley. Email: {\tt gvaliant@eecs.berkeley.edu}.} \and
Paul Valiant \thanks{MIT. Email: {\tt pvaliant@mit.edu}.}
\institute{}}
\date{\today}

\begin{document}

\maketitle

\newcommand{\PPAD}{\mathbf{PPAD}}
\newcommand{\NP}{\mathbf{NP}}

\begin{abstract}
We consider the problem of computing Nash Equilibria of \emph{action-graph games} (AGGs).  AGGs, introduced by Bhat and Leyton-Brown, is a succinct representation of games that encapsulates both `local'dependencies as in graphical games, and partial indifference to other agents' identities as in anonymous games, which occur in many natural settings. This is achieved by specifying a graph on the set of actions, so that the payoff of an agent for selecting a strategy depends only on the number of agents playing each of the neighboring strategies in the action graph. We present a Polynomial Time Approximation Scheme for computing mixed Nash equilibria of AGGs with constant treewidth and a constant number of agent types (and an arbitrary number of strategies), together with hardness
results for the cases when either the treewidth or the number of agent types is unconstrained. In particular, we show that even if the action graph is a tree, but the number of agent-types
is unconstrained, it is NP--complete to decide the existence of a pure-strategy Nash equilibrium and PPAD-complete to compute a mixed Nash equilibrium (even an approximate one); similarly for symmetric AGGs (all agents belong to a single type), if we allow arbitrary treewidth.  These hardness results suggest that, in some sense, our PTAS is as strong of a positive result as one can expect.
\end{abstract}

\section{Introduction}~\label{section:Introduction}

What is the likely behavior of autonomous agents in a variety of competitive environments?  This question has been the motivation for much of economic theory.  Partly due to the increasing prevalence of vast online networks over which millions of individuals exchange information, goods, and services, and the corresponding increasing importance of understanding the dynamics of such interactions, the Computer Science community has joined in the effort of studying game-theoretic questions.

%Work from both Economics and Computer Science has led to the emergence of a variety of characterizations of `expected' behavior, perhaps none as simple, compelling, and well studied as Nash equilibria.

Computing equilibria in games and markets has been extensively studied in the Economics and Operations Research communities since the 1960's, see e.g. \cite{Lemke-Howson,Rosenmuller,Wilson,Scarf,Taal}. Computational tractability has been recently recognized as an important prerequisite for modeling competitive environments and measuring the plausibility of solution concepts in Economics: if finding an equilibrium is computationally intractable, should we believe that it naturally arises? And, is it plausible that markets converge to solutions of computationally intractable problems?  Probably not --- but if so, we should certainly know about it.

Computing Nash equilibria in games, even in the case of two players, has been recently shown to be an intractable problem; in particular, it was shown to be complete for the class of fixed point computation problems termed PPAD~\cite{DP_2006,CD_2006}. This result on the intractability of computing Nash equilibria has sparked considerable  effort to find efficient algorithms for approximating such equilibria, and has increased the importance of considering special classes of games for which Nash equilibria might be efficiently computable.

For two-player games the hardness of computing approximate equilibria persists even if the required approximation is inverse polynomial in the number of strategies of the game \cite{CDT06}; similarly, hardness persists in graphical games if the required approximation is inverse polynomial in the number of players \cite{DP_2006,CDT06}. The same hardness results apply to special cases of the problem, e.g. win-lose games, where the payoff values of the game are restricted to $\{0,1\}$ \cite{Val05}, sparse bimatrix games, where the number of non-zero entries of each row an column of the payoff matrices is a constant, and two-player symmetric games \cite{GaleKuhnTucker}. The emerging question of the research in this field is: { Is there a Polynomial Time Approximation Scheme (PTAS) for Computing Nash Equilibria?} And, { which special cases of the problem are computationally tractable?}

The zero-sum two-player case is well-known to be tractable by reduction to linear programming \cite{vonNeumanLP}. Tractability persists in the case of low-rank two-player games, in which the sum $A+B$ of the payoff matrices of the players, instead of being $0$, has fixed rank; in this case, a PTAS exists for finding mixed Nash equilibria \cite{KannanTheobald}. In $n$-player graphical games, a PTAS has been provided if the tree-width is $O(\log n)$ and the maximum degree is bounded \cite{DasPapEC06}; in the case of dense graphical games, a quasi-polynomial time approximation scheme exists \cite{DasPapExhaustive}.

An important line of research on tractable special cases has explored games with symmetries. Mutli-player symmetric games with about logarithmically few strategies per player can be solved exactly in polynomial time by reduction to the existential theory of reals \cite{PapRough}. For congestion games, a pure Nash equilibrium can be computed in polynomial time if the game is a symmetric network congestion game \cite{FabPapTalwar}, and an approximate pure Nash equilibrium if the game is symmetric but not necessarily a network game and the utilities satisfy a ``bounded-jump condition'' \cite{ChienSinclair}. Another important class of games for which computing an approximate equilibrium is tractable is the class of anonymous games, in which each player is different, but does not care about the identities of the other players, as it happens for example congestion games, certain auction settings, and social phenomena \cite{Blonski99}; a PTAS for anonymous games with a fixed number
 of strategies has been provided in \cite{DP:anonymous07,DasPapExhaustive}. For a thorough study of the problem of computing pure Nash equilibria in symmetric and anonymous games see \cite{BFH_2007}.

In this paper, we consider another special class of games,  {\em Action Graph Games} (AGGs), that were introduced by Bhot and Leyton-Brown \cite{B_LB_2004}.  AGGs is a fully general game representation that succinctly captures both `local' dependencies as in graphical games, as well as partial indifference to other agents' identities as in anonymous games.  Strategies are represented as nodes in a graph, called \emph{strategy graph},  and the utility of an agent for selecting a strategy-node depends on the number of other agents selecting each of the neighboring strategies.  The only attribute that distinguishes players is the set of strategies that each player is allowed to play.  In particular, all agents who play a given strategy get the same payoff.  A variety of natural games can be concisely represented as AGGs, and we refer the reader to~\cite{B_LB_2004,J_LB_2007} for further discussion.

In the remainder of this section, we discuss previous work on AGGs and summarize our results. In the end of the section, we provide definitions.

\subsection{Previous Work}

Action graph games were first defined by Bhat and Leyton-Brown~\cite{B_LB_2004} who considered the problem of computing Nash equilibria of these games.  In particular, they analyzed the complexity of computing the Jacobian of the payoff function---a computation that is, in practice, the bottleneck of the continuation method of computing a Nash equilibrium.  They considered this computation for both general AGGs and AGGs with a single player type (symmetric AGGs), and found that this computation is efficient in the latter case.  For pure Nash equilibria, Jiang and Leyton-Brown~\cite{J_LB_2007}
show that deciding the existence of such equilibria in AGGs is $\NP$-complete, even in the case of a single player type and bounded degree.  On the other hand, they provide a polynomial time algorithm for finding pure-Nash equilibria in AGGs with constant number of player types when the strategy graph has bounded tree-width.

\subsection{Our Results}
We examine, and largely resolve the computational complexity of computing Nash equilibria in action graph games. We give a polynomial algorithm for computing an $\epsilon$-Nash equilibrium
for AGGs with constant treewidth and degree and a constant number of agent
types (and arbitrarily many strategies), together with hardness
results for the cases when either the treewidth or the number of
agent types is unconstrained.  In particular, we show that even if
the strategy graph is a tree with bounded degree but the number of agent types
is unconstrained, it is NP--complete to decide the existence of a
pure-strategy Nash equilibrium and PPAD--complete to compute a mixed
Nash equilibrium; similarly for AGGs in which all agents are a
single type if we allow the strategy graph to have arbitrary treewidth.  These hardness results
suggest that, in some sense, our PTAS is as strong of a positive
result as one can expect.   While Bhat and Leyton-Brown studied heuristics for computing mixed
Nash equilibria \cite{B_LB_2004}, the authors know of no previous
complexity theoretic results concerning computing mixed Nash
equilibria for AGGs apart from the $\PPAD$-hardness result that follows from
them being a generalization of normal-form games.

\subsection{Definitions}~\label{subsection:Action Graph Games}

In this section we give a formal definition of AGGs and introduce the terminology that will be used in the remainder of this paper.  We follow the notation and terminology introduced in~\cite{J_LB_2007}. \\

\begin{definition}
An \emph{action-graph game}, $A$, is a tuple $\langle P,\bold{S},G,u
\rangle$ where
\begin{itemize}
     \item{$P:=\{1,\ldots, n\}$ is the set of agents.}
     \item{$\bold{S}:=(S_1,\ldots,S_n),$ where $S_i$ denotes the set of pure strategies that agent $i$ may play.}
     \item{For convenience, let $S := \bigcup_i S_i = \{s_1,\ldots,s_{|S|}\}$ denote the set of all strategies, and thus each $S_i \subseteq S$.
        Also, we write $S_i = \{s_{i, 1}, s_{i, 2}, \ldots, s_{i,
        |S|_i}\}$
        Furthermore, we'll let $s(i)$ denote the strategy played by agent $i$.}
     \item{For any $S' \subset S$, let $\Delta(S')$ denote the set of valid configurations of agents to strategies $s \in S'$; we represent a configuration $D(S') \in \Delta(S')$ as an $|S'|$-tuple $D(S')=\{n_1,\ldots,n_{|S'|}\}$ where $n_i$ is the number of agents playing the $i^{th}$ strategy of $S'$.}
     \item{$G$ is a directed graph with one node for each action $s_i$.  Let $\nu : S\rightarrow 2^S$, be the neighbor relation induced by graph $G$, where $s' \in \nu(s)$ if the edge $(s',s) \in G$.  Note that self-loops are allowed, and thus it is possible that $s \in \nu(s)$.  We refer to $G$ as the \emph{strategy graph} of $A$.}
     \item{The utility function $u$ assigns identical utilities to all agents playing a given strategy $s$, with the utility depending only on the number of agents playing neighboring strategies.  Formally, $u: \Delta(S) \rightarrow \mathbb{R}^{|S|}$, via maps $u_1,\ldots,u_{|S|}$ where $u_i: \Delta[\nu(s_i)] \rightarrow \mathbb{R}$ defines the common utility of all agents playing strategy $s_i$.}
\end{itemize}
\end{definition}

Note that AGGs are fully expressive because any games can be
written as an action graph game in which the strategy sets of different players are disjoint, and the strategy graph $G$ is complete.

We now define a further type of possible symmetry between agents
that will be important in our analysis of the complexity of
computing Nash equilibria.

\begin{definition}
We say that an AGG has $k$ \emph{player types} if there exists a
partition of the agents into $k$ sets $P_1,\ldots,P_k$, such that if
$p,p' \in P_i$, then $S_p = S_{p'}$.  (The terminology
of~\cite{J_LB_2007} refers to such games as $k$-symmetric AGGs.)
\end{definition}

Since agents who play the same
strategy receive the same utility, all agents of a given type are
identical---for example an AGG with a single player type is a
symmetric game.  While the number of player types does not
significantly alter the description size, decreasing the number of
player types constrains the space of possible Nash equilibria; this is the motivation for considering AGGs with few player types as a possible class of tractable games.

A \emph{strategy profile},
$M:=[m_1,\ldots, m_n]$, with $m_i =(p_{i,1},\ldots,p_{i,|S_i|})$
assigns to each agent a probability distribution over the possible
strategies that the agent may play, with $\Pr[
s(i)=s_{i,k}]=p_{i,k}$ where $s_k$ is the $k^{th}$ element of $S_i$.  Thus a given strategy profile induces an expected utility for each player $\E[u|M] = \sum_{D \in \Delta} u(D) \Pr(D),$ where the probability is with respect to the strategy profile $M$.

\begin{definition}
A strategy profile $M$ is a \emph{Nash-equilibrium} if no player
can increase her expected utility by changing her strategy $m_i$ given the strategy profiles $m_{-i}$ of the other agents. That is,
for all strategy profiles $m_i'$. $\E[u_i|m_{-i},m_i] \ge \E[u_i|m_{-i},m_i']$.
\end{definition}

\begin{definition}
A strategy $m\in M$ is an
$\epsilon$-\emph{Nash-equilibrium} if no player can increase her
expected utility by more than $\epsilon$ by changing her strategy profile.
\end{definition}

Note that there is the slightly stronger definition of an $\epsilon$--Nash equilibrium in which, for all agents $i$, the expected utility of playing every strategy $s$ in the support of $m_i$ is at most $\epsilon$ less than the expected utility of playing a different $s' \in S_i$.  We do not stress the distinction, as our PTAS finds such an $\epsilon$--Nash equilibrium, and our hardness results apply to the weaker definition given above. 

\section{PTAS}

%
%\begin{definition}
%An \emph{game}, $G$, is a tuple $\langle P,\bold{S},u \rangle$ where
%\begin{itemize}
%     \item{$P:=\{1,\ldots, n\}$ is the set of agents.}
%     \item{$\bold{S}:=(S_1,\ldots,S_n),$ where $S_i$ denotes the set of pure strategies that agent $i$ may play.}
%     \item{For convenience, we'll let $s(i)$ denote the strategy played by agent $i$.}
%     \item{The utility function $u = (u_1, \ldots, u_n)$ assigns identical utilities to all agents playing a given strategy $s$.  $u_i: \prod_{i' \in P} S_i \rightarrow
%     \mathbb{R}$.}
%\end{itemize}
%\end{definition}
%
%\begin{definition}  An \emph{game with $k$-player types} is a
%game $G = \langle P,\bold{S},u \rangle$ such that there exists a
%partition of the $n$ players into $k$ sets $P_1, P_2, \ldots, P_k$,
%such that for any strategy profile $s$,  $u_i(s) = u_j(\pi(s))$
%where $\pi$ is any a permutation on the players such that for all $i
%\in P$, if $i \in P_j$ if and only if $\pi(i) \in P_j$.
%\end{definition}
%
%Note that an action graph game has $k$-player types if there there
%exists a collection $\{S^i\}_{1 \leq i \leq k}$ where $S^i \subseteq
%S$ such that each players strategy space $S_i$ is equal to $S^j$ for
%some $j$.

Action graph games have properties of both anonymous games and
graphical games.  As such, one might expect that classes of AGGs
that resemble tractable classes of anonymous or graphical games
could have efficiently computable equilibria.  For anonymous games,
the symmetry imposed by the limited number of types implies the
existence of a highly symmetric mixed equilibrium which seems easier to
find than asymmetric equilibria.  For graphical games with small
treewidth, the tree structure allows for an efficient
message-passing dynamic-programming approach. For AGGs with a
bounded number of player types and a strategy graph of constant
treewidth, we give a PTAS for computing $\epsilon$-Nash equilibria
that uses both the symmetry implied by bounding the number of player
types, and a dynamic programming approach that exploits the tree
structure. While these conditions might seem strong, we show in
Section~\ref{section:Hardness Results} that if either condition is
omitted the problem of computing an $\epsilon$-Nash equilibrium is
hard.

\begin{theorem}~\label{thm:PTAS}
For any fixed constants $d$, $k$, and $t$, an AGG $A$ with $k$
player types and strategy graph $G_A$ with bounded degree $d$  and
treewidth $t$, an $\epsilon$-Nash equilibrium can be computed in time
polynomial in $|A|,1/\epsilon$.
\end{theorem}

We begin with a fact about games with few player types.

\begin{fact}~\cite{Nash51}~\label{fact:sym eq}
Any AGG with $k$ player types has
a Nash equilibrium where all players of a given type play identical mixed strategy profiles.  Formally, there is a strategy profile $M=[m_1,\ldots,m_n]$ such that if  $S_i=S_j$, then $m_i=m_j$.  We refer to such
equilibria as \emph{type-symmetric} equilibria.
\end{fact}

The the high-level outline of the PTAS is as follows: we discretize
the space of mixed strategy profiles such that each player may play
a given strategy with probability $N \delta$ for $N \in \N$, and
some fixed $\delta>0$ that will depend on $\epsilon$ and $n$.  We
also discretize the space of target expected utilities into the set
$V = \{0,\epsilon/2,\epsilon,\ldots,1\}$. Then, for each $i \in
\{0,\ldots,|V|\}$, starting from the leaves of the strategy-graph
tree, we employ  dynamic programming to efficiently search the
discretized strategy space for a type-symmetric $\epsilon$-Nash
equilibrium in which each strategy in the support has an expected
utility close to $v_i$.  To accomplish this we associate to each
strategy $s_i$ a polynomially sized table expressing the set of
probabilities with which $s_i$ could be played so that some
assignment of probabilities to the strategies below $s_i$ in the
strategy tree could be extended to such an $\epsilon$-Nash
equilibrium for the whole game.  The following lemma guarantees the
existence of such a type-symmetric $\epsilon$-Nash equilibrium.

\begin{lemma}~\label{lemma:symmetric equilibrium}
Given an $n$-player AGG $A$, with 1 player type and strategy graph
$G_A$ with maximum degree $d$, for any $\delta>0$ there is a
strategy profile $Q=(q_1,\ldots,q_{|S|})$ with each $q_i$ a multiple
of $\delta$ and the property that if all agents play profile $Q$,
for any strategy $s$ in the support of $Q$, $\E[u_s|Q] \ge
\E[u_{s'}|Q] - 2 \delta d n$ for all $s' \in S$.
\end{lemma}

The following standard fact will be necessary in our proof:

\begin{fact}~\label{fact:binomial}
For binomially distributed random variables
$X=B(n,p),Y=B(n,p+\delta)$ $$\max_{k} |\Pr(X=k)-\Pr(Y=k)| \le n
\delta.$$
\end{fact}

\begin{prevproof}{Lemma}{lemma:symmetric equilibrium}
From Fact~\ref{fact:sym eq}, there exists a strategy profile
$P=(p_1,\ldots,p_{|S|})$ which is a Nash equilibrium of $A$.
Consider a strategy profile $Q$ with the property that $q_i=0$ if
$p_i=0$, and otherwise $|q_i-p_i| \le \delta$.  (Note that such a
profile clearly exists.)\begin{comment} Such a profile can be
obtained from $P$ by, for example, iterating through all $i$ with
$p_i>0$ and either rounding up or down to obtain $q_i$, where the
rounding is chosen so as to maintain a maximum disparity of
$\delta$, since it sums to 1 blah \end{comment}  For a given
strategy $s_i$ with $\nu(s_i)=d$, we now show that $$\left|\E[u_s|Q] -
\E[u_s|P]\right| \le \delta d n,$$ from which our lemma follows.

For a single neighbor $s_j$ of $s_i$, from Fact~\ref{fact:binomial}
$\left|\Pr(D(s_j)=k|Q)-\Pr(D(s_j)=k|P)\right| \le \delta n$, and
thus if we were to replace $p_j$ by $q_j$ in profile $P$, this
change would affect the expectation of playing $s_i$ by at most
$\delta n $.  Applying this reasoning to each of the $d$ neighbors
completes our proof.
\end{prevproof}

We now describe the PTAS; for clarity we describe the algorithm in
the case that $k=1$, and $G_A$ is a tree with maximum degree 3,
although it extends easily to a constant number of player types and
constant treewidth.  The following definition simplifies our
description of the algorithm.

\begin{definition}~\label{def:partial eq}
We say that some set of strategy profiles is an
\emph{$\epsilon$-partial equilibrium} for a subset $S' \subset S$ of
strategies if, for all strategies $s_i$ played with nonzero
probability, the expected utility of playing $s_i$ is at most
$\epsilon$ less than the expected utility of playing some other $s'
\in S'$.
\end{definition}

Consider a fixed $\epsilon>0$, and an $n$-player AGG $A=\langle
P,\bold{S},G,u \rangle$ with $k$ types, with strategies
$S=\{s_1,\ldots,s_{|S|}\}$ and strategy graph $G_A$ that is a tree
with maximum degree of 3. Arbitrarily choose some strategy with
degree 1 as the root of $G_A$, and without loss of generality denote
it by $s_1$. Given a strategy $s_i$, let $s_{R(i)},  s_{L(i)}$
denote the right and left children of $s_i$ in the strategy graph.
If $s_i$ has only one child, let $s_{R(i)}$ denote this child and
$s_{L(i)}=null$. Fix $\delta = \frac{\epsilon}{2dn}$.   Set
$V:=\{v_0,\ldots,v_{|V|} \}$ where $v_i:=\frac{i \epsilon}{2}$.

Let $f_i$ represent a table of size
$\frac{1}{\delta^4}|V|=poly(\frac{1}{\epsilon},n,|S|)$ associated
with strategy $s_i$, which can be thought of as a function $f_i:
I_{\delta}^4 \times V \rightarrow \{0,1\}$ where
$I_{\delta}=\{0,\delta, 2\delta,\ldots,1\}$.  The function
$f_i(p,p_R,p_L,w,v_i)$ will indicate whether there is a
type-symmetric $\frac{\epsilon}{4}$-partial equilibrium in the set
of strategies below $s_i$ in $G_A$ with expected utility near $v_i$
and strategies $s_i,s_{R(i)},s_{L(i)}$ played with probabilities
$p,p_R,p_L$, respectively, where the probability of choosing
strategy $s_i$, or one that lies below $s_i$ is $w$. In addition to
$f_i$, we also construct another table $g_i:I_{\delta}^4 \times V
\rightarrow I_{\delta}^2$ that will facilitate the reconstruction of
an $\epsilon$-Nash equilibrium after having computed all the tables
$f_i$ and $g_i$.  The $g_i$ will record the total weight used in
each of the partial solutions which were combined.  A given element
of $g_i$ will never be used if the corresponding element of $f_i$ is
0, and for simplicity we neglect to define $g_i$ for these entries.
Starting from the leaves, we calculate the tables $f_i,g_i$ as
follows:
\begin{itemize}
    \item{if $s_i$ is a leaf, $f_i(p,p_R,p_L,w,v_i)=1$ iff $p=w$ and $p_R=p_L=0$}
    \item{if $s_i\neq s_1$ has one child, $f_i(p,p_R,p_L,w,v)=1$ and $g_i(p,p_R,p_L,w,v)=(w-p,0)$ iff $p_L=0$ and there exist $q_R,q_L,w' \in I_{\delta}$ such that the following conditions hold:
        \begin{itemize}
            \item{$f_{R(i)}(p_R,q_R,q_L,w',v)=1$}
            \item{if $p_R>0$ the expected utility of playing $s_{R(i)}$ is in $(v-\frac{\epsilon}{2},v+\frac{\epsilon}{2})$ given that $s_i,s_{R(R(i))},s_{L(R(i))}$ are played with respective probabilities $p,q_R,q_L$.}
            \item{if $p_R=0$ the expected utility of playing $s_{R(i)}$ is at most $v+\frac{\epsilon}{2}$ given that $s_i,s_{R(R(i))},s_{L(R(i))}$ are played with respective probabilities $p,q_R,q_L$.}
            \item{$w=w'+p$}
        \end{itemize}}
    \item{if $s_i\neq s_0$ has two children, $f_i(p,p_R,p_L,w,v)=1$ and $g_i(p,p_R,p_L,w,v)=(w^R,w^L)$ if there exist $q_R^R$,$q_L^R$,$w^R$,$q_R^L$\\$q_L^L$,$w^L \in I_{\delta}$ such that the following conditions hold:
        \begin{itemize}
            \item{$f_{R(i)}(p_R,q_R^R,q_L^R,w^R,v)=1=f_{L(i)}(p_L,q_R^L,q_L^L,w^L,v)$}
            \item{if $P_R>0$ the expected utility of playing $s_{R(i)}$ is in $(v-\frac{\epsilon}{2},v+\frac{\epsilon}{2})$ given that $s_i,s_{R(R(i))},s_{L(R(i))}$ are played with respective probabilities $p,q_R,q_L$.  Analogously for  the utility of playing $s_{L(i)}$ if $p_L>0$.}
            \item{if $p_R=0$ the expected utility of playing $s_{R(i)}$ is at most $v+\frac{\epsilon}{2}$ given that $s_i,s_{R(R(i))},s_{L(R(i))}$ are played with respective probabilities $p,q^R_R,q^R_L$.  Analogously for  the utility of playing $s_{L(i)}$ if $p_L=0$.}
            \item{$w=w^R+w^L+p$}
        \end{itemize}}
        Note that there may be multiple choices of $q_R^R,q_L^R$,$w^R,q_R^L$,$q_L^L,w^L$ that satisfy the above conditions, in which case $g_i(p,p_R,p_L,w,v)=(w^R,w^L)$ can be assigned to an arbitrary choice of such $w^R,w^L$.
    \item{set $f_1(p,p_R,p_L,w,v)=1$ iff $p_L=0$, $w=1$, and there exist $q_R,q_L,w' \in I_{\delta}$ such that the following conditions hold:
        \begin{itemize}
            \item{$f_{R(1)}(p_R,q_R,q_L,w',v)=1$}
            \item{if $p_R>0$ the expected utility of playing $s_{R(1)}$ is in $(v-\frac{\epsilon}{2},v+\frac{\epsilon}{2})$ given that $s_1,s_{R(R(1))},s_{L(R(1))}$ are played with respective probabilities $p,q_R,q_L$.}
            \item{if $p_R=0$ the expected utility of playing $s_{R(1)}$ is at most $v+\frac{\epsilon}{2}$ given that $s_1,s_{R(R(1))},s_{L(R(1))}$ are played with respective probabilities $p,q_R,q_L$.}
            \item{if $s_1>0$ the expected utility of playing $s_1$ is in $(v-\frac{\epsilon}{2},v+\frac{\epsilon}{2})$ given that $s_{R(1)}$ is played with probability $p_R$.}
            \item{if $s_1=0$ the expected utility of playing $s_1$ is at most $v+\frac{\epsilon}{2}$ given that $s_{R(1)}$ is played with probability $p_R$.}
            \item{$w=w'+p$}
        \end{itemize}}
\end{itemize}

The following lemma ensures that the tables $f_i$ behave as hoped,
and we find at least one approximate Nash equilibria.  The proof
follows from the definition above and induction on the tree
structure.  For the sake of brevity we omit a formal proof.

\begin{lemma}~\label{lemma:construction works}
   %If $f_i(p,p_R,p_L,w,v)=1$, then there is some strategy profile $(q_1,\ldots,q_{|S|})$ with $q_i=p$, $q_{R(i)}=p_R$, $q_{L(i)}=p_L,$  that is an $\epsilon$-partial equilibrium for the subset of strategies $S'$ that lies below $s_i$ in the tree $G_A$, such that $\sum_{i\in S' \cup \{s_i \}} q_i = w$.  Furthermore,
   Given a strategy profile $(q_1,\ldots,q_{|S|})$ that is a
   $\frac{\epsilon}{4}$-Nash equilibrium of the form guaranteed in
   Lemma~\ref{lemma:symmetric equilibrium}, then for all
   $i$, $f_i(q_i,q_{R(i)},q_{L(i)},w,v)=1$, where $v$ is chosen to be a multiple of $\epsilon/2$ and to be
   within $\epsilon/2$ of the expected utility of playing any strategy in the support, and $w$
   is the sum of the weights on strategy $i$ and its descendants in the tree.
\end{lemma}

The following two lemmas demonstrate that the tables $f_i$ can be
computed efficiently, and that given the tables, an $\epsilon$-Nash
equilibrium can be efficiently computed.

\begin{lemma}~\label{lemma:f computable}
The tables $f_0,\ldots,f_{|S|}$ and $g_0,\ldots,g_{|S|}$  can be
computed efficiently.
\end{lemma}
\begin{proof}
  The size of each table is polynomially sized.  $f_i$ and $g_i$
  can be computed efficiently given the tables $f_{R(i)}$, $f_{L(i)}$.
\end{proof}

\begin{lemma}~\label{lemma:given tables}
Given the tables $f_0,\ldots,f_{|S|}$, and $\epsilon$-Nash
equilibrium can be found efficiently.
\end{lemma}
\begin{proof}
    Starting from the root of the tree $G_A$, we will populate a
    strategy profile $(p_1,\ldots,p_{|S|})$ that will be a
    type-symmetric $\epsilon$-Nash equilibrium.
    Lemmas~\ref{lemma:construction works} and~\ref{lemma:symmetric equilibrium}
    guarantee that there will be some choice of $p,p_R\in I_{\delta}$,
    and $v \in V$ such that $f_1(p,p_R,0,1,v)=1$.  We set $p_1=p$, $p_{R(1)}=p_R$,
    and `pass' $w=1-p_1$ and $v$ to strategy $s_{R(1)}$.  For all other strategies
    $s_i$, with $i \ge 2$ such that $p_i$ has already been fixed but
    $p_{R(i)},p_{L(i)}$ have not been fixed yet, the parent of $s_i$
    will have passed a pair $w,v$.  In the case that $s_i$ has one child,
    from our construction of $f_i$, it follows that there must be a
    choice of $q_R \in I_{\delta}$ such that $f_i(p_i,q_R,0,w-p_i,v)=1$.
    We then set $p_{R(i))}=q_R$ and $p_{L(i)}=q_L$ and pass the pair
    $w'=w-p_i,v$  to $s_{R(i)}$.  In the case that $s_i$ has two children,
    from our construction of $f_i$, it follows that
    there must be a choice of  $q_R,q_L \in I_{\delta}$
    such that $f_i(p_i,q_R,q_L,w,v)=1$, and
    $g_i(p_i,q_R,q_L,w,v)=(w^R,w^L)$.  We then set
    $p_{R(i))}=q_R,$ $p_{L(i))}=q_L,$ and pass the pairs
    $w^R,v$ and $w^L,v$ to the right and left children, respectively.

    From our construction, it follows that $\sum_i p_i = 1$, and that
    for every strategy $s_i$ in the support, the expected
    utility of playing that strategy is at most $\epsilon$
    less than the expected utility of playing any other strategy.
    In particular, $(p_1,\ldots,p_{|S|})$ is an $\epsilon$-Nash equilibrium.

\vspace{10pt} This algorithm and proof easily extends to the case
where there are a constant $k$ player types: simply create
tables $f_i^{(j)}$ and $g_i^{(j)}$ for each type $j: 1 \leq j \leq
k$ and strategy $i: 1 \leq j \leq |S|$ and proceed analogously but
additionally require that $f_i^j(p_i,p_R,p_L,w^j,v^j) = 0$ if $i \not\in
S_j$ and $p_i \neq 0$.  That is, enforce that each player type only
play the strategies available.

\vspace{10pt} Additionally, using a standard technique, the algorithm extends to the case
where the tree-width is bounded by some constant $t$.  Intuitively, in this case the
strategy graph decomposes into a tree over cliques of size $t$ of vertices on
the graph.  All the vertices in each clique are processed
simultaneously.  Because $t$ is constant, the increase in running time is polynomial.  \GSnote{should we cite the "standardness"
here?}

\end{proof}

Finally, one could consider the setting in which there is an unbounded number of player types, but each type consists of a
connected region of the tree.  An analogue of the above algorithm
can handle this setting provided
not too many player types can play any particular strategy.

\begin{definition}  Let $S_1, \ldots, S_k$ be subset of a vertices
of a tree $T$.  Define $S^c_i$ to be smallest connected region of
$T$ such that $S_i \subseteq S^c_i$.  We define the overlap of $S_1,
\ldots, S_k$ of $T$ to be $max_{t \in T} |i : t \in S^c_i|$.
\end{definition}

\begin{corollary}
~\label{cor:PTAS} For any fixed constants $c$ and $d$, an AGG $A$
with $k$-player types $S_1, \ldots, S_k$, and strategy graph $G_A$
which is a tree with bounded degree 1, and the overlap of $S_1,
\ldots, S_k$ on $G_A$ is at most $c$, an $\epsilon$-Nash equilibrium
can be computed in time polynomial in $|A|,1/\epsilon$.
\end{corollary}

\begin{proof}  For each strategy $i \in S$, define tables $f_i^{(j)}$ and $g_i^{(j)}$ for type $j$ only when
$i \in S_j^c$ and proceed as in the case of $k$ player types.  \end{proof}

\section{Hardness Results}~\label{section:Hardness Results}

In this section we state and prove our four hardness results.  We
show that it is (1) NP--complete to decide the existence of pure-strategy Nash equilibria, and (2) PPAD complete to approximate general (mixed Nash) equilibria for the classes of action graph games that either (a) have action graphs of treewidth 1 or (b) are symmetric (all agents are of a single type).  Our two hardness results for pure equilibria will come from reductions from
the NP--complete problem CIRCUITSAT, and follow the approach
of~\cite{SV_2006}.  Our hardness results for approximating mixed Nash
equilibria are via equilibria-preserving gadgets that let us reduce from the
PPAD-complete problem of computing equilibria in the class of graphical games where the maximum degree is $3$ and each player has only two possible strategies.  We begin by showing that action graph games are in the class PPAD.

\subsection*{Mapping Action Graph Games to Graphical Games}
\label{sec:actiontographical}

We show the following result which reduces the problem of computing a Nash equilibrium of an action graph game to the problem of computing a Nash equilibrium of a graphical game. Since the latter is in PPAD \cite{PapParityArgument}, it follows that the former is in PPAD as well.
\begin{theorem} \label{th:generalTheoremForGames}
Any action-graph game $A$ can be mapped in polynomial time to a
graphical game $\G$ so that there is a polynomial-time computable
surjective mapping from the set of Nash equilibria of $\G$ to the set
of Nash equilibria of $A$.
%
\iffalse Moreover, the problem of computing a Nash equilibrium in an action-graph game is polynomial-time reducible to the problem of computing a Nash equilibrium of a 2-player game.\fi
\end{theorem}
\begin{proof}
Let us define a bounded division-free straight-line program to be an arithmetic
binary circuit with nodes performing addition, subtraction, or multiplication on their
inputs, or evaluating to pre-set constants, with the additional constraint that the values
of all the nodes remain in [0, 1].

We will show that there exists a bounded division-free staight-line program of polynomial size in the description of the action graph game which, given a mixed strategy profile $M:= \{(p_{i,1},\ldots,p_{i,|S_i|})\}_{i=1}^n$, computes, for every agent $i$, $i=1,\ldots,n$, and for every pure strategy $s_i$, $s_i \in S_i$, of that agent, the expected utility that this agent gets for playing pure strategy $s_i$. The proof then follows from Theorems 1 and 2 of \cite{DasFabPap:ICALP06}.

Without loss of generality, we will show that there exists a straight-line program of polynomial size for computing the expected utility of agent $1$ for playing pure strategy $s_{1,1}$. For this purpose, let $\mathcal{N}:=\nu(s_{1,1})$ and $\Delta:=\Delta(\nu(s_{1,1}))$. Also, for any subset $P' \subseteq P$ of the agents, let $\Delta_{P'}$ be the set of valid configurations of the agents of the set $P'$ to the strategies in $\mathcal{N}$, represented as $|\mathcal{N}|$-tuples of numbers; then, for every $D \in \Delta_{P'}$, let $\Pr_{P'}[D]$ be the probability that configuration $D$ arises from the set of agents $P'$, where the measure $\Pr_{P'}$ is taken over the mixed strategies of agents in $P'$. Using this notation, the expected payoff of agent $1$ for playing $s_{1,1}$ can be written as follows

\begin{align}
\mathcal{U}_{1,s_{1,1}}:=\sum_{D \in \Delta_{P\setminus\{1\}}} u_{s_{1,1}}(D + 1_{s_{1,1}}) \Pr{}_{P\setminus \{1\}}[D], \label{eq:costis1}
\end{align}
where $1_{s_{1,1}}$ is an $|\mathcal{N}|$-tuple of numbers having a $1$ at the coordinate corresponding to strategy $s_{1,1}$ and $0$ everywhere else, and where $D + 1_{s_{1,1}}$ represents coordinate-wise addition.

From Equation \eqref{eq:costis1}, it follows that, if there is a bounded division-free straight-line program of polynomial size which computes the values $\{\Pr{}_{P\setminus\{1\}}[D]\}_{D \in \Delta_{P\setminus\{1\}}}$, a straight-line program for computing $\mathcal{U}_{1,s_{1,1}}$ can be constructed at an additional cost of $O(|\Delta_{P\setminus\{1\}}|) = O(|A|)$ arithmetic gates. To conclude the proof, we prove the following lemma.

%Hence, it is enough to argue that $\Pr{}_{P_{-1}}[D]$ can be computed for a specific value of $D \in \Delta_{-1}$.

\begin{lemma}
Let $P_j=\{j+1,\ldots,n\} \subseteq P$ and suppose that there exists a bounded division-free staight-line program computing the values $\{\Pr_{P_j}[D]\}_{D \in \Delta_{P_j}}$. Also, suppose that the size of this straight-line program is bounded by $g$. Then there exists a straight-line program of size bounded by $g+O(|A|\cdot(|A|+|S|))$ which computes the values $\{\Pr_{P_j \cup \{j\}}[D]\}_{D \in \Delta_{P_j \cup \{j\}}}$.
\end{lemma}

\begin{proof}
Note first that $|\Delta_{P_j \cup \{j\}}| = O(|A|)$, where $|A|$ is the description size of the action graph game. Then, for every $D\in\Delta_{P_j \cup \{j\}}$, it holds that
\begin{align}
\Pr{}_{P_j \cup \{j\}} [D] &= \sum_{s_{j,k} \in S_j \setminus \mathcal{N}}{\Pr{}_{\{j\}}[\text{$j$ plays $s_{j,k}$}] \Pr{}_{P_j}[D]}\notag\\&~~~~~~~~~~~~~~~~~~~~ +\sum_{\begin{minipage}{2.5cm}\centering $s_{j,k} \in S_j \cap \mathcal{N}$,\\ $D' \in \Delta_{P_j}$:\\ $1_{s_{j,k}}+D' = D$ \end{minipage}}{ \Pr{}_{\{j\}}[1_{s_{j,k}}] \Pr{}_{P_j}[D']}, \label{eq:costis2}
\end{align}
where in the above expression $\Pr_{P_j}[D]=0$, if $D \notin \Delta_{P_j}$. Observe that in \eqref{eq:costis2} the first summation has at most $|S|$ terms and the second at most $|A|$ terms. Hence, given the values $\{\Pr_{P_j}[D]\}_{D \in \Delta_{P_j}}$, the above expression can be evaluated with $O(|A|+|S|)$ arithmetic operations. The result follows.
\end{proof}

From the lemma above it follows that there exists a straight-line program of size bounded by $O(n\cdot|A|\cdot(|A|+|S|)) = O(|A|^3)$ for computing $\{\Pr{}_{P\setminus\{1\}}[D]\}_{D \in \Delta_{P\setminus\{1\}}}$. This concludes the proof of the theorem, since all the intermediate values of the computations described above are in $[0,1]$.
\end{proof}

%\end{document}

\subsection*{A Copy Gadget}

As a preliminary to the hardness results of the next two subsections, we describe a \emph{copy gadget} which will prove useful
in both NP-completeness and PPAD-completeness results.  Intuitively, to simulate games $G$ of high treewidth by treewidth 1 action graph games $H$, we create several ``copies'' of each player, but only one copy of each edge relating players, thus ending up with a very ``sparse'' simulation, whose treewidth we can control.  Explicitly, given an AGG
$A$, and an agent $i$ whose strategy set consists of the two strategies $S_i=\{f_i,t_i\}$, our copy gadget will add two additional players $a,c$, of which player $c$ will be the ``copy'' and player $a$ is an auxiliary player, whose inclusion will allow player $i$'s strategies to be disconnected from player $c$'s.  We add strategies for $a$ and $c$ that are $\{f_a,t_a\}$ and $\{f_c,t_c\}$ respectively, and set the incentives so that in any Nash equilibrium $\Pr[
s(i)=t_i] = \Pr[s(b)=t_b]$ (and $\Pr[ s(i)=f_i] = \Pr[s(b)=f_b]$).

\begin{figure}
\begin{center}
\epsfig{file=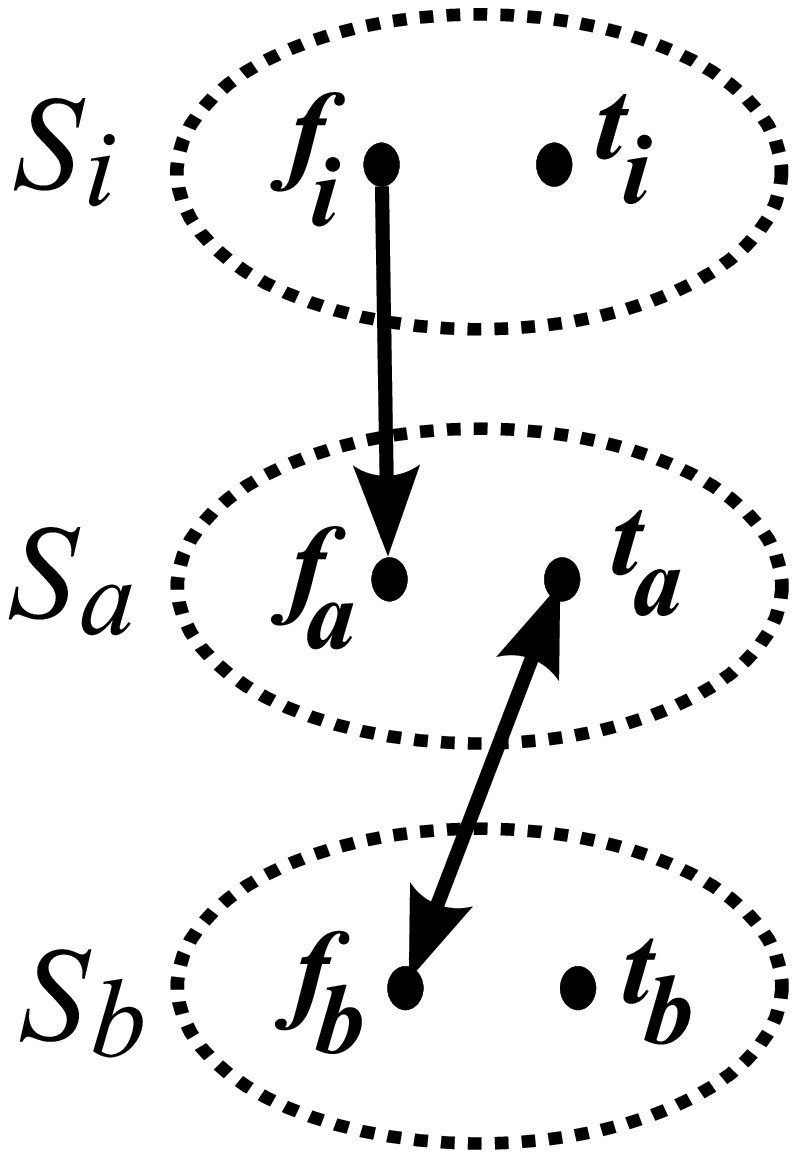,width=.4\textwidth} \caption{The
\emph{copy gadget}---in any Nash equilibrium
$\Pr[s(b)=f_b]=\Pr[s(i)=f_i]$.}\label{fig:copy gadget}
\end{center}
\end{figure}

\begin{definition}~\label{def:copy gadget}
Given an AGG $A=\langle P,\bold{S},G,u \rangle$, and an agent $i$
with two strategy choices $S_i=\{f_i,t_i\}$, we create AGG
$A'=\langle P',\bold{S'},G',u' \rangle$ from $A$ via the addition of
a \emph{copy gadget on $i$} as follows:
\begin{itemize}
    \item{$P':=P \cup \{a,c\}.$}
    \item{$\bold{S'}:=(S_1,\ldots,S_{|P|},S_a,S_c)$, where $S_a=\{f_a,t_a\}$, and $S_c=\{f_c,t_c\}$, where $f_a,t_a,f_c,t_c \not \in S$.}
    \item{$G'$ consists of the graph $G$ with the additional vertices corresponding to $f_a,t_a,f_c,t_c$, and the directed edges $(f_i,f_a), (t_a,f_c),(f_c,t_a)$.}
    \item{$u'$ is identical to $u$ for all strategies in $S'\backslash \{S_a \cup S_c\}$, and for a configuration $D$, $u'(f_a)=D(f_i),$ $u'(t_a)=1-D(f_c),$ and $u'(f_c)=1-2 D(t_a).$}
\end{itemize}
\end{definition}
See Figure~\ref{fig:copy gadget} for a depiction of the copy gadget.

\begin{lemma}~\label{lemma:copy gadget}
Given an AGG $A$ and an agent $i$, the addition of a copy gadget on
$i$ yields $A'$ that satisfies the following properties:
\begin{itemize}
    \item{The description size of $A'$ is at most a constant larger than $A$.}
\iffalse    \item{For every $\epsilon$--Nash equilibrium $M'$ of $A'$, the restriction of $M'$ to agents $1,\ldots,n_A$ is a Nash equilibrium of $A$, and for any equilibrium $M$ of $A$, there exist distributions $M_a:=(p_{a,f},p_{a,t}), M_b:=(p_{b,f},p_{b,t})$ over strategies for agents $a$ and $b$ such that $M$ together with $M_a,M_b$ is an equilibrium of $A'$.}\fi
    \item{In the strategy graph $G_{A'}$, $f_c$ and $t_c$ are not path connected to either $f_i$ or $t_i$.}
    \item{In every $\epsilon^2$--Nash equilibrium with agent $i$'s profile $(p_{i,f},1-p_{i,f})$, agent $c$'s profile will have $|p_{c,f}-p_{i,f}|\leq \epsilon$ (and $|p_{b,t}-p_{i,t}|\leq \epsilon$).}
\end{itemize}
\end{lemma}

\begin{proof}
The first two properties follow directly from
Definition~\ref{def:copy gadget}.  For the third property, assume
otherwise and consider the case where $p_{c,f} > \epsilon+p_{i,f}$.  Agent
$a$'s expected utility for playing $f_a$ is $p_{i,f}$, and is
$p_{c,f}$ for playing $t_a$, thus our assumption that $p_{c,f} >
\epsilon+p_{i,f}$ implies that agent $a$ must be playing $t_a$ with probability at least $1-\epsilon$ since the game
is at $\epsilon^2$--equilibrium.  Given that $a$ plays $t_a$ with probability at least $1-\epsilon$, agent $b$ maximizes
her utility by playing $t_c$, and thus $p_{c,f}\leq \epsilon$, which contradicts our assumption that $p_{c,f}$ is the larger of $p_{c,f},p_{i,f}$, namely at least $\frac{1}{2}$.  An analogous argument applies to rule out the case $p_{c,f} < p_{i,f}-\epsilon$.
\end{proof}

\subsection{PPAD-Completeness}

Our PPAD-Completeness results are reductions from the problem of
computing equilibria in graphical games, and rely on the following
fact due to~\cite{CDT06}.

\begin{fact}~\label{fact:PPAD}
For the class of graphical games with $n$ players, maximum degree $3$ and
payoffs in $\{0,1,2\}$, it is PPAD--complete to compute $\epsilon$-Nash equilibria where
$\epsilon \propto 1/poly(n)$.
\end{fact}

\begin{theorem}~\label{thm:PPAD tw1}
Computing a Nash equilibrium for AGGs with strategy graph $G_A$ is
PPAD-complete even if \\$treewidth(G_A)=1$, and $G_A$ has constant degree.
\end{theorem}

\begin{figure}
\begin{center}
\epsfig{file=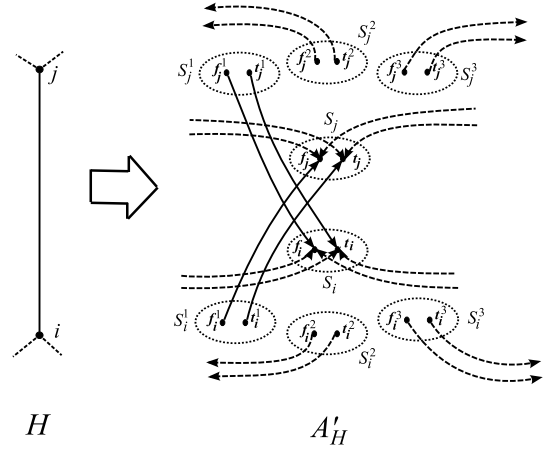,width=.6\textwidth} \caption{The
transformation from the graphical game $H$ to the AGG $A_H'$.  For
simplicity, the internal strategies and edges associated with the
copy gadgets are omitted.}~\label{fig:H to A_H}
\end{center}
\end{figure}

\begin{proof}
From Theorem \ref{th:generalTheoremForGames} this problem is in PPAD.

To show PPAD-hardness, we reduce from the known PPAD-hard problem of Fact \ref{fact:PPAD}.  Given an instance of such a graphical game $H$, we construct an AGG $A_H'$ with
treewidth $1$ and maximum degree 4 with similar description size to
$H$ such that there a polynomial time mapping from $\epsilon$--Nash equilibria of $A_H'$ to the $\epsilon$--Nash equilibria of $H$.  We construct $A_H'$ via the
intermediate step of constructing an AGG $A_H$ which will be
equivalent to $H$ and might have large treewidth.  From $A_H$, we
construct $A_H'$ using our copy gadget to reduce the treewidth of
the associated strategy graph.  See Figure~\ref{fig:H to A_H} for a
depiction of the reduction.

The construction of $A_H$ is straightforward: for each player $i_H$
in the graphical game, we have a corresponding player $i_A$ in the
AGG with strategy set $S_{i_A}=\{f_i,t_i\}$, corresponding to the
two strategies that $i_H$ may play in $H$.  For each undirected edge between players
$(i,j) \in H$, we add directed edges between the $t$ nodes $(t_j,t_i),(t_i,t_j)$, and edges between the $f$ nodes $(f_j,f_i),(f_i,f_j)$ to the strategy graph
$G_{A_H}$ of $A_H$.  We define utilities $u$ by simulating the utility functions from the original game $H$: from each $f$ strategy connected to $i_A$ in the AGG we know that if it is played then the corresponding $t$ strategy is not played and vice versa; thus we have recovered the strategy choice of each neighbor of $I_H$ in original graphical game; we then apply the utility function of the graphical game to compute the utility in the AGG.  We do the symmetric procedure for the $t$ nodes of the AGG.  From the
construction, it is clear that $H$ and $A_H$ represent the same game
via the correspondence $i_H \rightarrow i_A$, and in particular an
$\epsilon$--Nash equilibrium of one game will correspond to an
$\epsilon$--Nash equilibrium of the other game via the natural
mapping.

We obtain $A_H'$ from $A_H$ by making three copies of each $i_A$ via
the copy gadget.  Thus for each $i$ there are agents $i_A,
i_A^1,i_A^2,i_A^3$ with $S_{i_A^k}=\{f_{i_A}^k,t_{i_A}^k\}$.
Finally, for each of the (at most three) outgoing edges of $f_{i_A}$
that are not part of copy gadgets, i.e the edges of the form
$(f_{i_A},f_{j_A})$, we replace the edge by $(f_{i_A}^k,f_{j_A})$,
with each $f_{i_A}^k$ having at most one outgoing edge, and modify
the utility function $u$ analogously so as to have the utility of
strategy $f_{j_A}$ depend on $f_{i_A}^k$ instead of $f_{i_A}$.
Analogous replacements are made for the outgoing edges of $t_{i_A}$.
Since the copied strategies $f_{i_A}^k,t_{i_A}^k$ are disconnected
from the original strategies $f_{i_A},t_{i_A}$ the longest path in
the strategy graph $G_{A_H'}$ associated with $A_H'$ has length at
most 4, with maximum degree 6, and $treewidth(G_{A_H})=1$.  (See
Figure~\ref{fig:H to A_H}.) Lemma~\ref{lemma:copy gadget} guarantees
that the transformation from $A_H$ to $A_H'$ increases the
representation size by at most a constant factor.  Further, from an $\frac{1}{144}\epsilon^2$--Nash equilibrium of $A_H'$ we can extract an $\epsilon$--Nash equilibrium of $A_H$ by simply ignoring the new players: all of the copies $i_A^k$ of a player $i_A$ will play strategies with probabilities within $\frac{1}{12}\epsilon$ of the probabilities of playing the original by Lemma~\ref{lemma:copy gadget}; thus the joint distribution of any triple of these will have joint distribution within $\frac{1}{4}\epsilon$ of the ``true'' joint distribution; since each utility has magnitude at most 2 the computed utilities will be within $\frac{1}{2}\epsilon$ of the utilities computed in $A_H$; thus each of the mixed strategies of a player $i_A$ in $A_H'$, interpreted as a strategy in $A_H$ will yield utility within $\epsilon$ of optimal.  From Fact \ref{fact:PPAD} we conclude that finding an $\frac{1}{144}\epsilon^2$--Nash equilibrium of $A_H'$ is PPAD complete for any polynomial $\epsilon$, yielding the desired result.
\end{proof}

We now turn our attention to AGGs that have a constrained number of
player types.

\begin{theorem}~\label{thm:PPAD sym}
Computing a Nash equilibrium for symmetric AGGs ($1$ player type) is
PPAD-complete even if the strategy graph $G_A$ has bounded degree.
\end{theorem}

To show PPAD-hardness, as above we reduce from the known PPAD-hard problem of
computing Nash equilibria in graphical games of degree at most $3$
where each player chooses between $2$ strategies $f,t$ and has
utility $0$, $1$, or $2$.  Given such a graphical game $H$, we will
reduce it to an AGG $A_H$ that has strategies $f_i,t_i$
corresponding to the two strategies that agent $i$ may choose in
$H$.  Intuitively, if our reduction is to be successful there are
several properties of $G_H$ that seem necessary.  First, in every
Nash equilibrium of $G_H$, there must be at least one agent playing
either $f_i$ or $t_i$ for every $i$.  This is accomplished by giving
agents a bonus payment if they choose either of the two strategies
of a sparsely-played $f_i,t_i$ pair.  Second, there must be some
unambiguous mapping between the number of agents playing $f_i$ and
$t_i$ in $A_H$ to a choice of actions of agent $i$ in $H$.  This is
accomplished via the MAJORITY function: if more agents play $f_i$
than $t_i$ in $A_H$, we say that $i$ plays $f$.  This motivating intuition is formalized in the
proof below.

\begin{prevproof}{Theorem}{thm:PPAD sym}
From Theorem \ref{th:generalTheoremForGames} this problem is in PPAD

To show PPAD-hardness, we reduce from the known PPAD-hard problem of
computing Nash equilibria in graphical games of degree at most $3$
where each player chooses between $2$ strategies and has utility
$0$, $1$, or $2$.  Given an instance of such a graphical game $H$,
with $n$ agents, and some $\epsilon>0$ we construct the symmetric
AGG $A_H$ so that an $\epsilon$--Nash equilibrium of $A_H$ can be
efficiently mapped to a $2 \epsilon$--Nash equilibrium of $H$.  We
construct $A_H=\langle P,\bold{S},G_{A},u \rangle$ as follows:
\begin{itemize}
    \item{$P:=\{1,\ldots,3 c n\}$ with $c>\frac{64}{\epsilon^2}$.}
    \item{$\bold{S}:= (S,\ldots,S)$, that is, each player has identical (symmetric) strategy set $S:=\{f_1,t_1,\ldots,f_n,t_n\}$ where strategies $f_i$ and $t_i$ correspond to the two strategies of the $i^{th}$ agent of $H$.}
    \item{For every undirected edge $(i,j)$ in the graph of $H$, the strategy graph $G_A$ has the eight directed edges $(f_i,f_j)$, $(f_j,f_i)$, $(f_i,t_j)$, $(t_j,f_i)$, $(t_i,f_j)$, $(f_j,t_i)$, $(t_i,t_j)$, $(t_j,t_i).$  Furthermore, for all $i \in \{1,\ldots, n\}$, $G_A$ contains the edges $(f_i,t_i),(t_i,f_i)$ and the self loops $(f_i,f_i)$ and $(t_i,t_i)$.}
    \item{To simplify the description of the utility function $u$, it will be useful to define the indicator functions\\ $I_1[D(f_1,t_1)],\ldots,I_n[D(f_n,t_n)]$ where $$I_i[D(f_i,t_i)]:= \left\{ \begin{array}{cc} f & \text{if $D(f_i) \ge D(t_i)$} \nonumber \\ t & \text{if $D(f_i)< D(t_i)$} \nonumber \end{array} \right.$$ Let $u$ assign utility to $f_i$ as a function of $D(\nu(i))$, where $\nu(i)$ denotes $i$'s neighbors, by applying the utility function for agent $i$ from $H$ on the simulated actions of her neighbors $j_1,j_2,j_3$ evaluated as $I_{j_1}[D(f_{j_1},t_{j_1})],I_{j_2}[D(f_{j_2},t_{j_2})],$ and $I_{j_3}[D(f_{j_3},t_{j_3})]$, respectively.  Finally, if $D(f_i)+D(t_i) \leq c$, $u$ assigns an extra $100$ utility to strategies $f_i$ and $t_i$.}
\end{itemize}
Observe that the description size of $A_H$ is polynomial in $cn$,
and thus is polynomial in the description size of $H$.  From Fact~\ref{fact:PPAD} our theorem will follow if we show that any $\epsilon^2$--Nash equilibrium of $A_H$ can
be efficiently mapped to an $\epsilon$--Nash equilibrium of $H$.

Consider the map from mixed strategy profiles of $A_H$ to mixed
strategy profiles of $H$ given by $\phi: \bold{M_A} \rightarrow
\bold{M_H}$ that assigns $M_H=[ (p_{1,f},
1-p_{1,f}),\ldots,(p_{n,f}, 1-p_{n,f}) ]$ by setting
$p_{i,f}:=\Pr_{M_A} (I_i=f)$ where the probability is taken over the
distribution over $\Delta$ defined by $M_A$.  It is clear that the
map $\phi$ can be computed efficiently, as it essentially involves simply evaluating multinomial distributions on 4 outcomes.
%XXXXXXXXXXXXXXXXXXXXXXXXXXXXXXXX DO WE NEED TO ELABORATE?

Before showing that $\phi$ maps $\epsilon$--equilibria to
$2\epsilon$--equilibria we first show that the ``extra utility" of 100 correctly incentivizes a large number of players to play on each strategy pair. We observe that in any mixed strategy
profile there will be at least one agent, $j$, who has probability at most $1/3$ of receiving a payoff of at least $100$.  Since his payoff from the simulation of $H$ is at most 2, such an agent's
expected utility is at most $35+\frac{1}{3}$, and thus any Nash equilibrium
mixed strategy profile must satisfy $100 \Pr( D(f_i)+D(t_i) < c) <
36, \forall i \in \{1,\ldots,n\}.$  If this were not the case, then
agent $j$ could improve her expected utility to at least $36$ by
always choosing strategy $f_i$, a contradiction.  The above
inequality implies that $$\E\left[\max\left(D(f_i),D(t_i)\right)
\right] > \frac{c}{4}.$$

We now proceed with the proof of correctness of the map $\phi$.  Let
$M_A$ be an $\epsilon$--Nash equilibrium of $A_H$, and
$M_H=\phi(M_A)$.  Consider a player $i$ in the graphical game, and a strategy of his that he plays with probability at least $\epsilon$.  Without loss of generality let this strategy be $f_i$.  We show that his utility for playing $f_i$ is at least his utility for playing his other choice, $t_i$, minus $\epsilon$; taken together, these statements imply that $M_H$ is an $\epsilon$--Nash equilibrium of the graphical game, as desired.

Since $\E\left[\max\left(D(f_i),D(t_i)\right)
\right] > \frac{c}{4}$, we have that if $f_i$ is played with probability at least $\epsilon$, namely if $\Pr(D(f_i)\geq D(t_i))\geq \epsilon$ then (by Chernoff bounds) we must have $\E[D(f_i)]\geq\frac{c}{6}$.  This implies that for at least one of the $3cn$ players $j$ in $A_H$, his probability of playing $f_i$ is at least $\frac{1}{18n}$, which is at least $2\epsilon$.  Thus, since $M_A$ is, by assumption, an $\epsilon^2$--Nash equilibrium, we have that player $j$'s utility for playing $f_i$ is at most $\frac{1}{2}\epsilon$ below his utility for playing $t_i$, when the other players play from $M_A$.  Further, by construction, each of these two utilities are within $\frac{1}{4}\epsilon$ of the utilities for player $i$ in the graphical game to play $f_i,t_i$ respectively, when the other players play from $M_H$: this is because fixing player $j$'s move from $M_{A,j}$ to one of $f_i,t_i$ changes each $I_k$ by at most the probability of $j$ playing one of $(f_k,t_k)$ divided by the standard deviation of $\max\left(D(f_i),D(t_i)\right)$, which is at least the square root of its expectation, $\sqrt{\frac{c}{4}}=\frac{4}{\epsilon}$.  Thus by the triangle inequality we have that player $i$'s utility for playing $f_i$ is no more than $\epsilon$ worse than playing $t_i$.  Since our choice of $i$ was arbitrary and a corresponding argument applies to the $t$ strategies, we conclude that $M_H$ is an $\epsilon$--Nash equilibrium of $H$, as desired.
\end{prevproof}

\subsection{NP--Completeness}

Both of our NP--completeness results are reductions from the
NP-Complete problem CIRCUITSAT and follow an approach employed
in~\cite{SV_2006}.

\begin{fact}~\label{fact:NP-complete}
It is NP-complete to decide satisfiability for the class of circuits consisting of AND, OR, and NOT gates, with maximum degree $3$ (in-degree plus out-degree).
\end{fact}

In our reductions from CIRCUITSAT, given a circuit $C$, we construct
an AGG $A_C$ that computes $C$ in the sense that pure strategy Nash
equilibria of $A_C$ map to valid circuit evaluations.
To this game we add two agents that have a simple pure-strategy equilibrium if $C$
evaluates to $true$, but when $C$ evaluates to $false$ play \emph{pennies}---a simple game
that has no pure strategy Nash equilibria.  Thus the existence of a
pure strategy Nash equilibrium is equivalent to the satisfiability of
$C$.

\begin{theorem}~\label{thm:NP tw1}
Deciding the existence of a pure strategy Nash equilibrium for AGGs
with strategy graph $G_A$ is NP-complete even if $treewidth(G_A)=1$,
and $G_A$ has constant degree.
\end{theorem}

\begin{proof}
Membership in NP is clear.  To show hardness, given a circuit $C$,
we construct the associated AGG $A_C:=\langle P,\bold{S},G_{A},u
\rangle$ as follows:
\begin{itemize}
    \item{$P:=\{1,\ldots,n, p_1,p_2\}$, where $n$ is the number of gates in $C$, and the gate corresponding to player $n$ is the output gate.}
    \item{$\bold{S}:= \left((f_1,t_1),\ldots,(f_n,t_n), (f_{p_1},t_{p_1}), (f_{p_2},t_{p_2})\right)$.}
    \item{For every pair of gates $i,j$ for which the output of gate $i$ is an input to gate $j$, $G_A$ has the edges $(f_i,f_j),$ and $(f_i,t_j)$.  Furthermore, we add edges $(f_n,f_{p_1}), (f_n,t_{p_1}), (f_n,f_{p_2}), (f_n,t_{p_2})$, and the edges $(f_{p_1}, f_{p_2})$, $(f_{p_1},t_{p_2})$, $(f_{p_2},f_{p_1})$, $(f_{p_2},t_{p_1})$.}
    \item{The utility function $u$ is defined as follows: if agent $i$ corresponds to an input gate, than strategies $f_i,t_i$ both have utility 0.  For any other agent $i$ corresponding to a gate of $C$, the payoff of strategy $f_i$ is $1$ or $0$ according to whether $f_i$ is the correct output value of gate $i$ given the values corresponding to the strategies played by neighboring agents/strategies.  Similarly for the payoff for strategy $t_i$. If $D(f_n)=0$, then $f_{p_1}$ and $t_{p_1}$ have utility 0, otherwise the utility of $p_1$ is 1 if $D(f_{p_1})=D(f_{p_2})$, and is 0 otherwise.  The utility of $p_2$ is 1 if $D(f_{p_1}) \neq D(f_{p_2})$, and is 0 otherwise.}
\end{itemize}
From the construction it is clear that if $C$ is satisfiable, there
is a pure strategy profile for agents $1,\ldots,n$ with agent $n$
playing $t_n$, such that agents $1,\ldots,n$ can not improve their
utility by deviating from their strategies.  Furthermore, $p_1$ will
be indifferent between her strategies, and $p_2$ will play the
opposite of $p_1$; in particular, there will be a pure strategy Nash
equilibrium.  If $C$ is not satisfiable, then any pure strategy
profile that is an equilibrium for agents $1,\ldots, n$ will have
$D(f_n)=1$, and thus $p_1$ will be incentivized to agree with $p_2$,
and $p_2$ will be incentivized to disagree, and thus $A_C$ will
admit no pure strategy Nash equilibrium.

To complete the proof, note that we can apply the copy gadget to
each agent of $A_C$, as was done in the proof of
Theorem~\ref{thm:PPAD tw1} to yield the game $A_C'$ that has
strategy graph of treewidth 1, and a mapping from equilibria of $A_C'$ to equilibria of $A_C$.
\end{proof}

\begin{theorem}~\label{thm:NP sym}
Deciding the existence of a pure strategy Nash equilibrium for
symmetric AGGs ($1$ player type) is NP-complete even if the strategy
graph $G_A$ has bounded degree.
\end{theorem}

\begin{proof}
Membership in NP is clear; to show hardness we proceed as was done
in the proof of Theorem~\ref{thm:NP tw1}, and obtain AGG $A_C$ from
circuit $C$.  Now, we make $A_C$ symmetric by retaining the same number of agents, but allowing each of them to
pick any of the strategies.   We modify the strategy graph $G$ by
adding edges $(f_x,t_x),(t_x,f_x),(f_x,f_x),(t_x,t_x)$ for each player $x$ from $A_C$, and extend
the utility function $u$ so that if $D(f_x)+D(t_x) >1$ then strategies
$f_x$ and $t_x$ have utility $-1$.  Thus in any pure strategy Nash
equilibrium $D(f_x)+D(t_x)=1$, and the reasoning in the proof of
Theorem~\ref{thm:NP tw1} applies to complete our reduction.
\end{proof}

\section{Conclusions and Open Problems}
We have presented results that further the understanding of the
computational complexity of computing approximate Nash equilibria in
action graph games.  We provided a polynomial-time algorithm for finding
approximate Nash equilibrium in action graph games under the restrictions that 1) the strategy graph of the game
has constant degree and constant treewidth and 2) the players can by classified by a constant number of types.
We showed that restricting the tree-width and the number of types of players
is necessary to avoid the problem becoming $\PPAD$-hard. Whether or
not the restriction on the degree of the graph is necessary remains
an open problem for future study.

We further showed the our algorithm extends to the case where the
underlying graph is a bounded-degree tree and there are an
arbitrary number of player types, but 1)  each player type is a
connected region of the graph and 2)  each strategy is available to
only a constant number of types.  It remains an open problem if the
second restriction is required.  Furthermore, perhaps there are
other restricted classes of games that circumvent the hardness result while
retaining some of the motivating features of general action graph games.

We also study the complexity of
computing pure Nash equilibria in action graph games with
$k$-types. While Jiang and Brown in \cite{J_LB_2007} show the
problem is tractable provided 1) $k$ is constant and 2) the graph
has bounded tree width, we show that both of these restrictions are
necessary.  Without either of them, finding pure Nash equilibrium in
action graph games becomes $NP$-complete.

\bibliographystyle{plain}
\bibliography{rbp_short}

\begin{thebibliography}{10}

\bibitem{B_LB_2004}
Navin A.~R. Bhat and Kevin Leyton-Brown.
\newblock Computing nash equilibria of action-graph games.
\newblock In {\em UAI}, pages 35--42, 2004.

\bibitem{Blonski99}
Matthias Blonski.
\newblock Anonymous games with binary actions.
\newblock {\em Games and Economic Behavior}, 28(2):171--180, August 1999.
\newblock available at
  http://ideas.repec.org/a/eee/gamebe/v28y1999i2p171-180.html.

\bibitem{CD_2006}
Xi~Chen and Xiaotie Deng.
\newblock Settling the complexity of two-player nash equilibrium.
\newblock In {\em FOCS '06: Proceedings of the 47th Annual IEEE Symposium on
  Foundations of Computer Science (FOCS'06)}, pages 261--272, Washington, DC,
  USA, 2006. IEEE Computer Society.

\bibitem{CDT06}
Xi~Chen, Xiaotie Deng, and Shang-Hua Teng.
\newblock Computing nash equilibria: Approximation and smoothed complexity.
\newblock In {\em FOCS}, pages 603--612, 2006.

\bibitem{ChienSinclair}
Steve Chien and Alistair Sinclair.
\newblock Convergence to approximate nash equilibria in congestion games.
\newblock In {\em SODA}, pages 169--178, 2007.

\bibitem{Val05}
Bruno Codenotti and Daniel \v{S}tefankovi\v{c}.
\newblock On the computational complexity of nash equilibria for (0, 1)
  bimatrix games.
\newblock {\em Inf. Process. Lett.}, 94(3):145--150, 2005.

\bibitem{DasFabPap:ICALP06}
Constantinos Daskalakis, Alex Fabrikant, and Christos~H. Papadimitriou.
\newblock The game world is flat: The complexity of nash equilibria in succinct
  games.
\newblock In {\em ICALP (1)}, pages 513--524, 2006.

\bibitem{DP_2006}
Constantinos Daskalakis, Paul~W. Goldberg, and Christos~H. Papadimitriou.
\newblock The complexity of computing a nash equilibrium.
\newblock In {\em STOC}, 2006.

\bibitem{DasPapEC06}
Constantinos Daskalakis and Christos~H. Papadimitriou.
\newblock Computing pure nash equilibria in graphical games via markov random
  fields.
\newblock In {\em ACM Conference on Electronic Commerce}, pages 91--99, 2006.

\bibitem{DP:anonymous07}
Constantinos Daskalakis and Christos~H. Papadimitriou.
\newblock Computing equilibria in anonymous games.
\newblock In {\em FOCS}, 2007.

\bibitem{DasPapExhaustive}
Constantinos Daskalakis and Christos~H. Papadimitriou.
\newblock On the exhaustive method for nash equilibria.
\newblock Manuscript, 2007.

\bibitem{BFH_2007}
F.~Fischer F.~Brandt and M.~Holzer.
\newblock Equilibria of graphical games with symmetries.
\newblock Technical Report TR07-136, Electronic Colloquium on Computational
  Complexity (ECCC), December 2007.

\bibitem{FabPapTalwar}
Alex Fabrikant, Christos~H. Papadimitriou, and Kunal Talwar.
\newblock The complexity of pure nash equilibria.
\newblock In {\em STOC}, pages 604--612, 2004.

\bibitem{GaleKuhnTucker}
D~Gale, HW~Kuhn, and AW~Tucker.
\newblock On symmetric games.
\newblock {\em Contributions to the Theory Games, Annals of Mathematics
  Studies}, 24, 1950.

\bibitem{J_LB_2007}
Albert~Xin Jiang and Kevin Leyton-Brown.
\newblock Computing pure nash equilibria in symmetric action graph games.
\newblock In {\em AAAI}, pages 79--85. AAAI Press, 2007.

\bibitem{KannanTheobald}
Ravi Kannan and Thorsten Theobald.
\newblock Games of fixed rank: a hierarchy of bimatrix games.
\newblock In {\em SODA}, pages 1124--1132, 2007.

\bibitem{Lemke-Howson}
C.~E. Lemke and Jr~J.~T.~Howson.
\newblock Equilibrium points of bimatrix games.
\newblock {\em SIAM Journal of Applied Mathematics}, 12:413--423, 1964.

\bibitem{Nash51}
John Nash.
\newblock Non-cooperative games.
\newblock {\em Annals of Mathematics}, 54(2):286--295, 1951.

\bibitem{PapParityArgument}
Christos~H. Papadimitriou.
\newblock On the complexity of the parity argument and other inefficient proofs
  of existence.
\newblock {\em J. Comput. Syst. Sci.}, 48(3):498--532, 1994.

\bibitem{PapRough}
Christos~H. Papadimitriou and Shmuel Safra.
\newblock The complexity of low-distortion embeddings between point sets.
\newblock In {\em SODA}, pages 112--118, 2005.

\bibitem{Rosenmuller}
J.~Rosenmuller.
\newblock On a generalization of the lemke--howson algorithm to noncooperative
  n-person games.
\newblock {\em SIAM Journal of Applied Mathematics}, 21:73--79, 1971.

\bibitem{Scarf}
H.~E. Scarf.
\newblock The approximation of fixed points of a continuous mapping.
\newblock {\em SIAM Journal of Applied Mathematics}, 15:1328--1343, 1967.

\bibitem{SV_2006}
Grant Schoenebeck and Salil Vadhan.
\newblock The computational complexity of nash equilibria in concisely
  represented games.
\newblock In {\em EC '06: Proceedings of the 7th ACM conference on Electronic
  commerce}, pages 270--279, New York, NY, USA, 2006. ACM.

\bibitem{Taal}
G.~van~der Laan and A.~J.~J. Talman.
\newblock On the computation of fixed points in the product space of unit
  simplices and an application to noncooperative n person games.
\newblock {\em Mathematics of Operations Research}, 7, 1982.

\bibitem{vonNeumanLP}
J.~von Neumann and O.~Morgenstern.
\newblock On the computation of fixed points in the product space of unit
  simplices and an application to noncooperative n person games.
\newblock {\em Theory of Games and Economic Behavior}, 1944.

\bibitem{Wilson}
R.~Wilson.
\newblock Computing equilibria of n-person games.
\newblock {\em SIAM Journal of Applied Mathematics}, 21:80--87, 1971.

\end{thebibliography}

\end{document}